\documentclass[aps,prl,twocolumn,superscriptaddress,showpacs,preprintnumbers,amsmath,amssymb]{revtex4}

\usepackage{graphicx}
\usepackage{dcolumn}
\usepackage{bm}

\def\bz{{B^0}}
\def\bzb{{\overline{B}{}^0}}

\def\kl{K_L^0}
\def\dE{{\Delta E}}
\def\mb{{M_{\rm bc}}}

\def\Dt{\Delta t}

\def\fol{f_{\rm ol}}

\newcommand{\sinbb}{{\sin2\phi_1}}
\def\sinbbeff{{\sin2\phi_1^{\rm eff}}}

\newcommand{\fCP}{f_{CP}}
\def\fcp{\fCP}
\newcommand{\ftag}{f_{\rm tag}}

\newcommand{\cala}{{\cal A}_f}
\newcommand{\cals}{{\cal S}_f}

\newcommand{\dm}{\Delta m_d}
\newcommand{\dmd}{\dm}
\def\taubz{{\tau_\bz}}

\def\ks{{K_S^0}}
\newcommand{\btosqq}{b \to s\overline{q}q}

\def\btoccs{b \to c\overline{c}s}
\newcommand*{\dwl}{\ensuremath{{\Delta w_l}}}
\newcommand*{\fq}{\ensuremath{q}}
\def\kz{{K^0}}
\def\kp{{K^+}}
\def\km{{K^-}}

\def\fzero{{f_0(980)}}

\def\pip{{\pi^+}}
\def\pim{{\pi^-}}
\def\piz{{\pi^0}}

\def\kl{{K_L^0}}
\def\bbar{{\overline{B}}}
\def\ufs{{\Upsilon(4S)}}

\def\jpsi{{J/\psi}}

\def\rhoz{{\rho^0}}

\def\lsig{{\cal L}_{\rm sig}}
\def\lbkg{{\cal L}_{\rm bkg}}
\def\rsigbkg{{\cal R}_{\rm s/b}}

\def\pbstar{p_B^*}

\def\etap{{\eta'}}

\def\kspm{{K_S^{+-}}}
\def\kszz{{K_S^{00}}}

\def\rsig{{R_{\rm sig}(\Dt)}}

\def\nbb{535}

\def\lint{492}

\def\lintsvdone{140}

\def\efftot{0.29\pm 0.01}

   \def\Nsigphiks{307 \pm 21}

   \def\Nsigphikl{114\pm 17}
   \def\Nsigetapks{1421 \pm 46}

   \def\Nsigetapkl{454 \pm 39}
   \def\Nsigksksks{185 \pm 17}
   \def\Nsigjpsiks{7484 \pm 87} 

   \def\Nsigjpsikl{6512 \pm 123}
\def\SphikzVal{+0.50}     \def\SphikzStat{0.21}     \def\SphikzSyst{0.06}
\def\sinbbphikzVal{\SphikzVal} 
\def\sinbbphikzStat{\SphikzStat} 
\def\sinbbphikzSyst{\SphikzSyst}
\def\AphikzVal{+0.07}     \def\AphikzStat{0.15}     \def\AphikzSyst{0.05}
\def\SphiksVal{+0.50}     \def\SphiksStat{0.23}     
 
\def\AphiksVal{+0.11}     \def\AphiksStat{0.16}     
\def\SphiklVal{-0.46}    \def\SphiklStat{0.56}
 
\def\AphiklVal{-0.15}     \def\AphiklStat{0.38}  
\def\SetapkzVal{+0.64}  \def\SetapkzStat{0.10}  \def\SetapkzSyst{0.04}
\def\sinbbetapkzVal{\SetapkzVal}  \def\sinbbetapkzStat{\SetapkzStat}  \def\sinbbetapkzSyst{\SetapkzSyst}
\def\AetapkzVal{-0.01}  \def\AetapkzStat{0.07}  \def\AetapkzSyst{0.05}
\def\SetapksVal{+0.67}  \def\SetapksStat{0.11} 
  
\def\AetapksVal{-0.03}  \def\AetapksStat{0.07}
 \def\mSetapklStat{0.24}
\def\SetapklVal{-0.46} \def\SetapklStat{\mSetapklStat}
 
\def\AetapklVal{+0.09}  \def\AetapklStat{0.16}
  \def\SksksksStat{0.32}  \def\SksksksSyst{0.08}
\def\sinbbksksksVal{+0.30}  
\def\sinbbksksksStat{\SksksksStat}  
\def\sinbbksksksSyst{\SksksksSyst}
\def\AksksksVal{+0.31}  \def\AksksksStat{0.20}  \def\AksksksSyst{0.07}
\def\SjpsikzVal{+0.642}  \def\SjpsikzStat{0.031}  \def\SjpsikzSyst{0.017}
\def\sinbbjpsikzVal{\SjpsikzVal}  
\def\sinbbjpsikzStat{\SjpsikzStat}  
\def\sinbbjpsikzSyst{\SjpsikzSyst}
\def\AjpsikzVal{+0.018}  \def\AjpsikzStat{0.021}  \def\AjpsikzSyst{0.014}
\def\SjpsiksVal{+0.643}  \def\SjpsiksStat{0.038}

\def\AjpsiksVal{-0.001}  
\def\AjpsiksStat{0.028}  

\def\mSjpsiklStat{0.057} 
\def\SjpsiklVal{-0.641} 
\def\SjpsiklStat{\mSjpsiklStat}

\def\AjpsiklVal{+0.045}  
\def\AjpsiklStat{0.033}  
\def\sinbbphikzResult{\sinbbphikzVal\pm\sinbbphikzStat\pm\sinbbphikzSyst}
\def\sinbbphikzResultSS
  {\sinbbphikzVal\pm\sinbbphikzStat\mbox{(stat)}\pm\sinbbphikzSyst\mbox{(syst)}}
\def\AphikzResult{\AphikzVal\pm\AphikzStat\pm\AphikzSyst}

\def\SphiksResultStat{\SphiksVal\pm\SphiksStat}

\def\AphiksResultStat{\AphiksVal\pm\AphiksStat}

\def\SphiklResultStat{\SphiklVal\pm\SphiklStat}

\def\AphiklResultStat{\AphiklVal\pm\AphiklStat}
\def\sinbbetapkzResult{\sinbbetapkzVal\pm\sinbbetapkzStat\pm\sinbbetapkzSyst}

\def\sinbbetapkzResultSS
  {\sinbbetapkzVal\pm\sinbbetapkzStat\mbox{(stat)}\pm\sinbbetapkzSyst\mbox{(syst)}}
\def\AetapkzResult{\AetapkzVal\pm\AetapkzStat\pm\AetapkzSyst}

\def\SetapksResultStat{\SetapksVal\pm\SetapksStat}

\def\AetapksResultStat{\AetapksVal\pm\AetapksStat}

\def\SetapklResultStat{\SetapklVal\pm\SetapklStat}

\def\AetapklResultStat{\AetapklVal\pm\AetapklStat}
\def\sinbbksksksResult{\sinbbksksksVal\pm\sinbbksksksStat\pm\sinbbksksksSyst}

\def\sinbbksksksResultSS
  {\sinbbksksksVal\pm\sinbbksksksStat\mbox{(stat)}\pm\sinbbksksksSyst\mbox{(syst)}}
\def\AksksksResult{\AksksksVal\pm\AksksksStat\pm\AksksksSyst}

\def\sinbbjpsikzResult{\sinbbjpsikzVal\pm\sinbbjpsikzStat\pm\sinbbjpsikzSyst}

\def\SjpsikzResultSS
  {\SjpsikzVal\pm\SjpsikzStat\mbox{(stat)}\pm\SjpsikzSyst\mbox{(syst)}}

\def\AjpsikzResult{\AjpsikzVal\pm\AjpsikzStat\pm\AjpsikzSyst}

\def\SjpsiksResultStat{\SjpsiksVal\pm\SjpsiksStat}

\def\AjpsiksResultStat{\AjpsiksVal\pm\AjpsiksStat}

\def\SjpsiklResultStat{\SjpsiklVal\pm\SjpsiklStat}

\def\AjpsiklResultStat{\AjpsiklVal\pm\AjpsiklStat}
\def\SphikzSystResol{0.04}

\def\SphikzSystBG{0.04}

\def\AphikzSystResol{0.01}

\def\AphikzSystBG{0.03}

\def\AphikzSystTagIntrfr{0.03}
\def\SetapkzSystResol{0.03}

\def\SetapkzSystBG{0.02}

\def\AetapkzSystVtx{0.02}

\def\AetapkzSystResol{0.02}

\def\AetapkzSystBG{0.02}

\def\AetapkzSystTagIntrfr{0.02}
\def\SksksksSystResol{0.05}

\def\SksksksSystBG{0.06}

\def\AksksksSystResol{0.02}

\def\AksksksSystBG{0.05}

\def\AksksksSystTagIntrfr{0.04}
\def\SjpsikzSystVtx{0.012}
\def\SjpsikzSystFbtg{0.004}
\def\SjpsikzSystResol{0.006}

\def\SjpsikzSystFit{0.007}
\def\SjpsikzSystBG{0.006}

\def\SjpsikzSystTagIntrfr{0.001}
\def\AjpsikzSystVtx{0.009}
\def\AjpsikzSystFbtg{0.003}
\def\AjpsikzSystResol{0.001}

\def\AjpsikzSystFit{0.004}
\def\AjpsikzSystBG{0.002}

\def\AjpsikzSystTagIntrfr{0.009}
\def\SIGMAetapkzAS{5.6}

\def\fKKKsBGinphiKs{2.75\pm 0.14}

\def\figWidth{0.23}

\begin{document}


\preprint{
  Belle Preprint 2006-28
}
\preprint{
  KEK Preprint 2006-40
}

\title{\boldmath 
      Observation of Time-Dependent $CP$
       Violation in $\bz\to\etap\kz$ Decays 
       and Improved Measurements of $CP$ Asymmetries in $\bz \to \phi\kz$, $\ks\ks\ks$
       and $\bz \to\jpsi\kz$ Decays
}
\date{\today}


\affiliation{Budker Institute of Nuclear Physics, Novosibirsk}
\affiliation{Chiba University, Chiba}
\affiliation{Chonnam National University, Kwangju}
\affiliation{University of Cincinnati, Cincinnati, Ohio 45221}
\affiliation{The Graduate University for Advanced Studies, Hayama, Japan} 
\affiliation{Gyeongsang National University, Chinju}
\affiliation{University of Hawaii, Honolulu, Hawaii 96822}
\affiliation{High Energy Accelerator Research Organization (KEK), Tsukuba}
\affiliation{Hiroshima Institute of Technology, Hiroshima}
\affiliation{University of Illinois at Urbana-Champaign, Urbana, Illinois 61801}
\affiliation{Institute of High Energy Physics, Chinese Academy of Sciences, Beijing}
\affiliation{Institute of High Energy Physics, Vienna}
\affiliation{Institute of High Energy Physics, Protvino}
\affiliation{Institute for Theoretical and Experimental Physics, Moscow}
\affiliation{J. Stefan Institute, Ljubljana}
\affiliation{Kanagawa University, Yokohama}
\affiliation{Korea University, Seoul}
\affiliation{Kyungpook National University, Taegu}
\affiliation{Swiss Federal Institute of Technology of Lausanne, EPFL, Lausanne}
\affiliation{University of Ljubljana, Ljubljana}
\affiliation{University of Maribor, Maribor}
\affiliation{University of Melbourne, Victoria}
\affiliation{Nagoya University, Nagoya}
\affiliation{Nara Women's University, Nara}
\affiliation{National Central University, Chung-li}
\affiliation{National United University, Miao Li}
\affiliation{Department of Physics, National Taiwan University, Taipei}
\affiliation{H. Niewodniczanski Institute of Nuclear Physics, Krakow}
\affiliation{Nippon Dental University, Niigata}
\affiliation{Niigata University, Niigata}
\affiliation{University of Nova Gorica, Nova Gorica}
\affiliation{Osaka City University, Osaka}
\affiliation{Osaka University, Osaka}
\affiliation{Panjab University, Chandigarh}
\affiliation{Peking University, Beijing}
\affiliation{Princeton University, Princeton, New Jersey 08544}
\affiliation{RIKEN BNL Research Center, Upton, New York 11973}
\affiliation{Saga University, Saga}
\affiliation{University of Science and Technology of China, Hefei}
\affiliation{Seoul National University, Seoul}
\affiliation{Shinshu University, Nagano}
\affiliation{Sungkyunkwan University, Suwon}
\affiliation{University of Sydney, Sydney NSW}
\affiliation{Tata Institute of Fundamental Research, Bombay}
\affiliation{Toho University, Funabashi}
\affiliation{Tohoku Gakuin University, Tagajo}
\affiliation{Tohoku University, Sendai}
\affiliation{Department of Physics, University of Tokyo, Tokyo}
\affiliation{Tokyo Institute of Technology, Tokyo}
\affiliation{Tokyo Metropolitan University, Tokyo}
\affiliation{Tokyo University of Agriculture and Technology, Tokyo}
\affiliation{Virginia Polytechnic Institute and State University, Blacksburg, Virginia 24061}
\affiliation{Yonsei University, Seoul}
  \author{K.-F.~Chen}\affiliation{Department of Physics, National Taiwan University, Taipei} 
  \author{K.~Hara}\affiliation{Nagoya University, Nagoya} 
  \author{M.~Hazumi}\affiliation{High Energy Accelerator Research Organization (KEK), Tsukuba} 
  \author{T.~Higuchi}\affiliation{High Energy Accelerator Research Organization (KEK), Tsukuba} 
  \author{K.~Miyabayashi}\affiliation{Nara Women's University, Nara} 
  \author{Y.~Nakahama}\affiliation{Department of Physics, University of Tokyo, Tokyo} 
  \author{K.~Sumisawa}\affiliation{High Energy Accelerator Research Organization (KEK), Tsukuba} 
  \author{O.~Tajima}\affiliation{High Energy Accelerator Research Organization (KEK), Tsukuba} 
  \author{Y.~Ushiroda}\affiliation{High Energy Accelerator Research Organization (KEK), Tsukuba} 
  \author{Y.~Yusa}\affiliation{Virginia Polytechnic Institute and State University, Blacksburg, Virginia 24061} 
% ...
\collaboration{The Belle Collaboration}

\begin{abstract}
  We present improved measurements of $CP$-violation parameters in
  $\bz\to$ $\phi\kz$, $\eta'\kz$, $\ks\ks\ks$ decays 
  based on a sample of $\nbb\times 10^6$ $B\bbar$ pairs collected at 
  the $\ufs$ resonance with the Belle detector at the KEKB 
  energy-asymmetric $e^+e^-$ collider.
  We obtain $\sinbbeff = \sinbbetapkzResultSS$ for $\bz\to\etap\kz$,
  $\sinbbphikzResultSS$ for $\bz\to\phi\kz$, and
  $\sinbbksksksResultSS$ for $\bz\to\ks\ks\ks$ decays.
  We have observed $CP$ violation in the $\bz\to\etap\kz$ decay
  with a significance of $\SIGMAetapkzAS$ standard deviations.
  We also perform an improved measurement of $CP$ asymmetries in
  $\bz\to\jpsi\kz$ decays and obtain $\sinbb = \SjpsikzResultSS$.
\end{abstract}

\pacs{11.30.Er, 12.15.Hh, 13.25.Hw}

\maketitle


Particles from physics beyond the standard model (SM) may contribute
to $\bz$ meson decays mediated by flavor-changing $b \to s$ transitions
via additional quantum loop diagrams, and potentially induce large deviations from
the SM expectation for time-dependent $CP$ asymmetries~\cite{bib:lucy}.  
In the decay chain $\Upsilon(4S)\to \bz\bzb \to f_{CP}f_{\rm tag}$,
where one of the $B$ mesons decays at time $t_{CP}$ to a 
$CP$ eigenstate $f_{CP}$ 
and the other decays at time $t_{\rm tag}$ to a final state  
$f_{\rm tag}$ that distinguishes between $B^0$ and $\bzb$, 
the decay rate has a time dependence~\cite{bib:sanda}
given by
\begin{eqnarray}
\label{eq:psig}
{\cal P}(\Delta{t}) = 
\frac{e^{-|\Delta{t}|/{\taubz}}}{4{\taubz}}
\biggl\{1 &+& \fq\cdot 
\Bigl[ \cals\sin(\dmd\Delta{t}) \nonumber \\
   &+& \cala\cos(\dmd\Delta{t})
\Bigr]
\biggr\}.
\end{eqnarray}
Here $\cals$ and $\cala$ are $CP$-violation parameters,
$\taubz$ is the $B^0$ lifetime, $\dmd$ is the mass difference 
between the two $B^0$ mass
eigenstates, $\Delta{t}$ = $t_{CP}$ $-$ $t_{\rm tag}$, and
the $b$-flavor charge $\fq = +1$ ($-1$) when the tagging $B$ meson
is a $B^0$ ($\bzb$).
In the SM, $CP$ violation arises only from the irreducible 
Kobayashi-Maskawa phase~\cite{Kobayashi:1973fv}
in the weak-interaction quark-mixing matrix~\cite{Cabibbo}.
The SM approximately predicts $\cals = -\xi_f\sinbb$\cite{footnote:phi1} and
$\cala =0$ for both $\btoccs$ and $\btosqq$ transitions,
where $\xi_f = +1$ $(-1)$ corresponds to  $CP$-even (-odd) final states,
in the leading order.
Recent SM calculations~\cite{bib:b2s_SM_uncertainties}
for the effective $\sinbb$ values, $\sinbbeff$, obtained from
$\bz\to\phi\kz$, $\etap\kz$ and $\ks\ks\ks$ agree with $\sinbb$,
as measured in $\bz\to\jpsi\kz$ decays, at the level of $0.01$. 
Thus comparison of measurements of $\cals$ and $\cala$ between modes 
is an important test of the SM. 

Previous measurements of $CP$ asymmetries in $\btosqq$ transitions
by Belle~\cite{Chen:2005dr}
and BaBar~\cite{bib:BaBar_sss}
differed from the SM expectation, 
although the deviation was not statistically significant.
BaBar has since updated their results~\cite{bib:BaBar_new}.
In this Letter we describe improved measurements of $\cals$ and
$\cala$ in
$\bz\to$
$\phi\ks$,
$\phi\kl$, 
$\eta'\ks$
and
$\ks\ks\ks$ decays using a data sample of $\lint$ fb$^{-1}$
($\nbb\times 10^6$ $B\bbar$ pairs), which is nearly twice that used for our
previous measurements. The analysis has also been improved by the
addition of the following decay chains:
$\bz\to\phi\ks$, $\phi\to\ks\kl$;
$\bz\to\eta'\kl$;
$\bz\to\eta'\ks$, $\ks\to\piz\piz$.
We also describe improved measurements of $\sinbb$ and $\cala$ in
$\bz\to\jpsi\ks$ and $\jpsi\kl$ decays using the same data sample;
our previous measurement used a $\lintsvdone$ fb$^{-1}$ data sample~\cite{bib:BELLE-CONF-0436}.
These modes have the largest statistics coupled with the smallest
theoretical uncertainties and thus provide a firm reference point for the SM.


At the KEKB energy-asymmetric 
$e^+e^-$ (3.5~GeV on 8.0~GeV) collider~\cite{bib:KEKB},
the $\Upsilon(4S)$ is produced
with a Lorentz boost of $\beta\gamma=0.425$ nearly along
the electron beamline ($z$).
Since the $B^0$ and $\bzb$ mesons are approximately at 
rest in the $\Upsilon(4S)$ center-of-mass system (cms),
$\Delta t$ can be determined from the displacement in $z$ 
between the $f_{CP}$ and $f_{\rm tag}$ decay vertices:
$\Delta t \simeq (z_{CP} - z_{\rm tag})/(\beta\gamma c)
 \equiv \Delta z/(\beta\gamma c)$.

The Belle detector~\cite{Belle} is a large-solid-angle magnetic
spectrometer that
consists of a silicon vertex detector (SVD),
a 50-layer central drift chamber (CDC), an array of
aerogel threshold Cherenkov counters (ACC),
a barrel-like arrangement of time-of-flight
scintillation counters (TOF), and an electromagnetic calorimeter
comprised of CsI(Tl) crystals (ECL) located inside
a superconducting solenoid coil that provides a 1.5~T
magnetic field.  An iron flux-return located outside of
the coil is instrumented to detect $K_L^0$ mesons and to identify
muons (KLM).



Charged tracks reconstructed with the CDC,
except for tracks from $\ks\to\pip\pim$ decays,
are required to originate from the interaction point (IP).
We distinguish charged kaons from pions based on
a kaon (pion) likelihood $\mathcal{L}_{K(\pi)}$
derived from the TOF, ACC, and $dE/dx$ measurements in the CDC.
Photons are identified as isolated ECL clusters
that are not matched to any charged track.
Candidate $\kl$ mesons are selected
from ECL and/or KLM hit patterns that are consistent 
with the presence of a shower induced by a $\kl$ meson.

The intermediate meson states are reconstructed from the following decays:
$\piz\to\gamma\gamma$,
$\ks \to \pip\pim$ (denoted by $\kspm$)
or $\piz\piz$ (denoted by $\kszz$),
$\eta\to\gamma\gamma$ or $\pip\pim\piz$,
$\rhoz\to\pip\pim$,
$\etap\to\rhoz\gamma$ or $\eta\pip\pim$,
$\phi\to \kp\km$
and $\jpsi\to\ell^+\ell^-$ ($\ell=\mu,e$).
We use all combinations of the intermediate states
with the exception of 
$\{\eta\to\pip\pim\piz,
\etap\to\rho\gamma,
\kszz\}$ 
candidates for
$\bz\to\{\etap\kszz,
\etap\kl,
\jpsi\ks
\}$ decays, respectively.
The reconstruction and selection criteria for $\bz\to\fCP$ decay candidates
are almost the same as in our previous measument~\cite{Chen:2005dr,bib:b2s_lp05}.
The $\kl$ and $\kszz$ candidates for $\eta'\kz$ are reconstructed with the same
method as used for $\phi\kz$~\cite{bib:b2s_lp05}.
We reconstruct the $\bz\to\ks\ks\ks$ decay
in the $\kspm\kspm\kspm$ or $\kspm\kspm\kszz$ final states.
In addition, $\phi\to\kspm\kl$ decays are used for
the $\bz\to\phi\kspm$ sample.
We identify candidate $\bz\to\fCP$ decays without a $\kl$ meson
using the energy difference $\dE\equiv E_B^*-E_{\rm beam}^{*}$ and
the beam-energy constrained mass $\mb\equiv\sqrt{(E_{\rm beam}^{*})^2-
(\pbstar)^2}$, where $E_{\rm beam}^{*}$ is the beam energy, and
$E_B^{*}$ and $\pbstar$ are the energy and momentum, respectively, of the
reconstructed $B$ candidate, all measured in the cms frame.
The signal candidates are selected by requiring 
$\mb\in(5.27,5.29)~{\rm GeV}/c^2$ 
and a mode-dependent $\dE$ window.
Only $\mb$ is used to identify the decay $\bz\to\phi\ks$ followed by
$\phi\to\ks\kl$.
Other candidate $\bz\to\fcp$ decays with a $\kl$ are selected by
requiring $\pbstar\in(0.2,0.45)~{\rm GeV}/c$ for $\bz\to\jpsi\kl$ candidates and
$\pbstar\in(0.2,0.5)~{\rm GeV}/c$ for the others.

The dominant background for the $\btosqq$ signal comes from
continuum events $e^+e^- \rightarrow q\overline{q}$ where $q={u,d,s,c}$.
To distinguish these topologically jet-like events from the spherical $B$ decay signal 
events, we combine a set of variables~\cite{Chen:2005dr}
that characterize the event topology
into a signal (background) likelihood variable $\cal L_{\rm sig}$ 
($\cal L_{\rm bkg}$),
and impose loose mode-dependent requirements on the likelihood ratio
$\rsigbkg \equiv \lsig/(\lsig+\lbkg)$.

The contributions from $B\overline{B}$ events 
to the background for $\bz\to\fcp$ candidates with a $\kl$ are
estimated with Monte Carlo (MC) simulated events.
The small ($\sim 3\%$) $B\overline{B}$ combinatorial background
in $\bz\to\etap\ks (\eta'\to\rhoz\gamma)$ is estimated using MC events.
We reject $\ks\ks\ks$ candidates 
if one of the $\ks$ pairs has an invariant mass within
$\pm 2 \sigma$ of the $\chi_{c0}$ mass or $\pm 1 \sigma$ of the $D^0$ mass, 
where $\sigma$ is the $\ks\ks$ mass resolution.
The fraction of $\bz\to\kp\km\ks$ and $\bz\to\fzero\ks~(\fzero\to\kp\km)$ events 
in the $\bz\to\phi\ks$ sample is estimated to be $\fKKKsBGinphiKs$\% and zero 
within error, respectively, from the Dalitz plot for $B\to\kp\km K$
candidates~\cite{Garmash:2003er}. 

The $b$-flavor of the accompanying $B$ meson is identified
from inclusive properties of particles
that are not associated with the reconstructed $\bz \to \fCP$ 
decay. 
The tagging information is represented by two parameters, the $b$-flavor
charge $\fq$ and $r$~\cite{bib:fbtg_nim}. 
The parameter $r$ is an event-by-event,
MC-determined flavor-tagging dilution factor
that ranges from $r=0$ for no flavor
discrimination to $r=1$ for unambiguous flavor assignment.
For events with $r > 0.1$, the wrong tag fractions for six $r$ intervals,
$w_l~(l=1,6)$, and their differences 
between $\bz$ and $\bzb$ decays, $\dwl$,
are determined
using semileptonic and hadronic $b\to c$
decays~\cite{Chen:2005dr,bib:BELLE-CONF-0436}. 
If $r \le 0.1$,
we set the wrong tag fraction to 0.5, and
therefore the tagging information is not provided.
The total effective tagging efficiency is determined to
be $\efftot$.

The vertex position for the $\fCP$ decay 
is reconstructed using charged tracks that have
enough SVD hits~\cite{bib:resol_nim}.
The $\ftag$ vertex is obtained with well-reconstructed 
tracks that are not assigned to $\fCP$.
A constraint on the interaction-region profile 
in the plane perpendicular to the beam axis
is also used with the selected tracks.
\begin{figure}
\includegraphics[width=0.48\textwidth,clip]{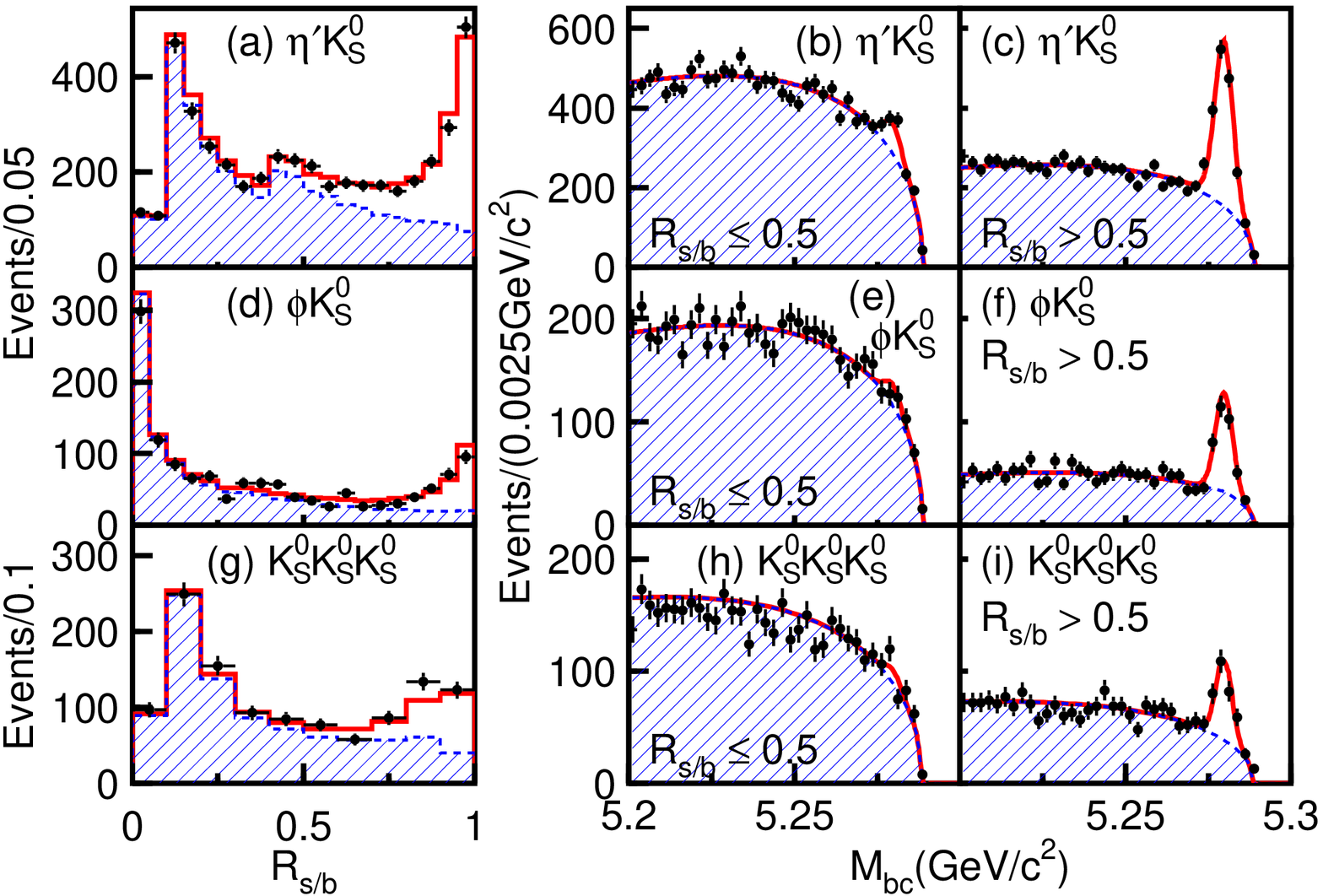}
\includegraphics[width=0.45\textwidth,clip]{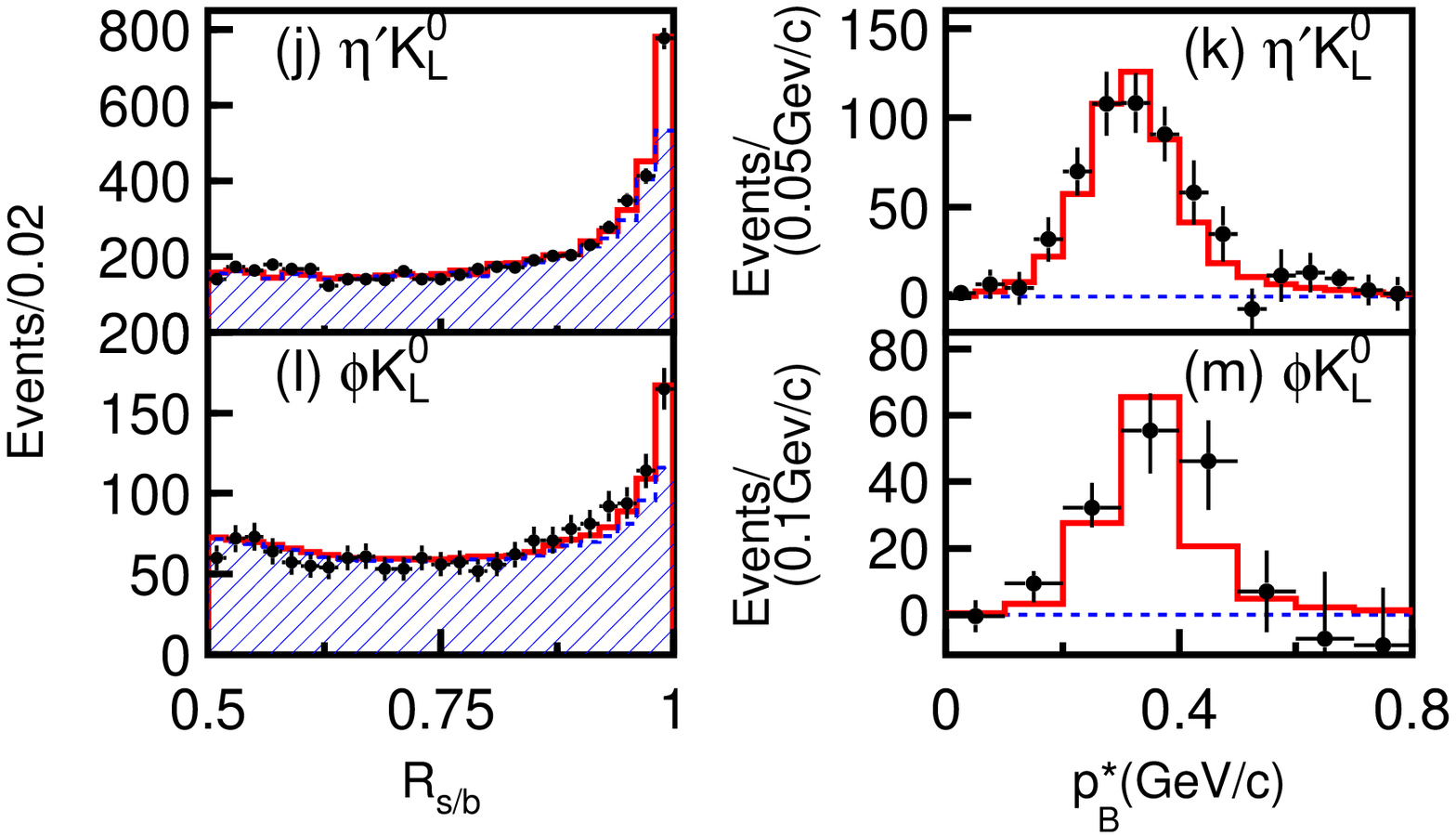}
\caption{
$\rsigbkg$, $\mb$ and $\pbstar$ distributions for reconstructed candidates:
$\rsigbkg$, $\mb$ with $\rsigbkg \le 0.5$ and $\mb$ with $\rsigbkg > 0.5$
distributions for (a, b and c) $\bz\to\etap\ks$, (d, e and f) $\bz\to\phi\ks$,
and (g, h and i) $\bz\to\ks\ks\ks$;
$\rsigbkg$ and $\pbstar$ for (j and k) $\bz\to\etap\kl$ and (l and m)
$\bz\to\phi\kl$.
The solid curves and histograms show the fits to signal plus background distributions,
and hatched areas show the background contributions.
Background contributions are subtracted in figures (k) and (m).
}
\label{fig:rec}
\end{figure}
%
%
\begin{figure}
\includegraphics[width=0.22\textwidth]{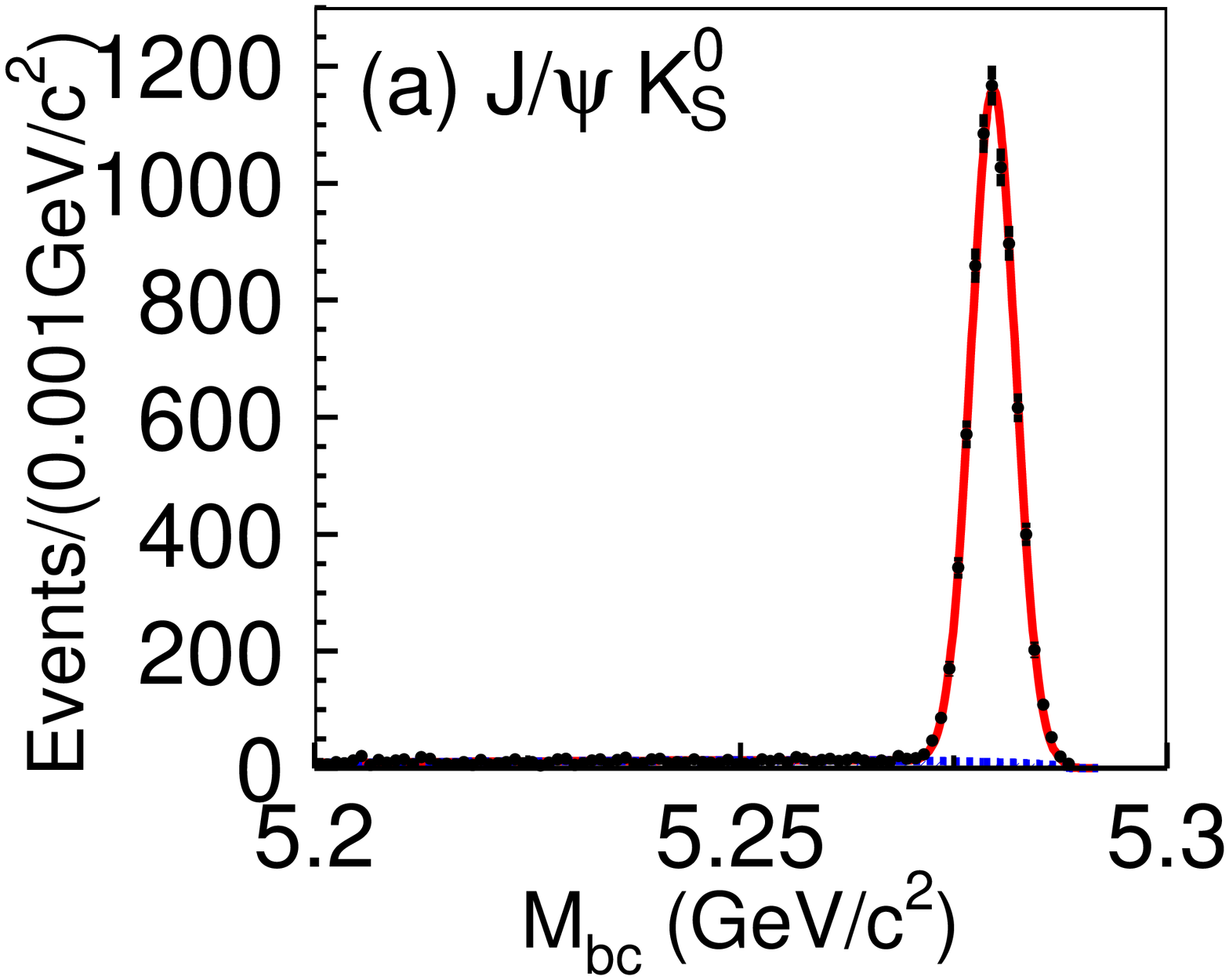}
\includegraphics[width=0.22\textwidth]{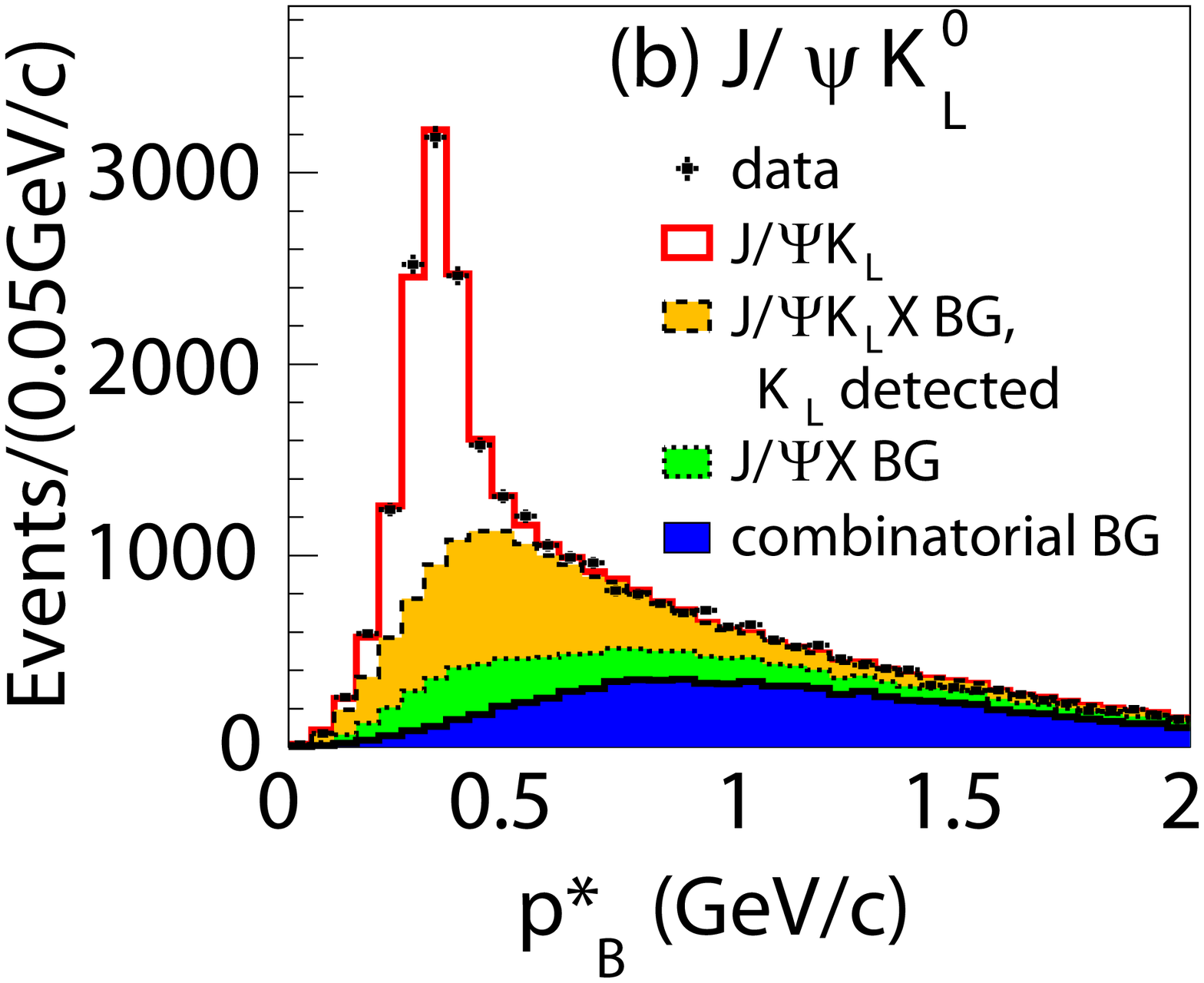}
\caption{(a) $\mb$ distribution 
in the $\dE$ signal region for selected $\bz\to\jpsi\ks$ candidates
and (b) $\pbstar$ distribution for selected $\bz\to\jpsi\kl$ candidates.
}
\label{fig:recjpsi}
\end{figure}

Figures~\ref{fig:rec}(a)-(m) show the distributions of reconstructed variables
$\rsigbkg$, $\mb$ and $\pbstar$
after flavor tagging and vertex reconstruction
for $\bz\to\etap\kz$, $\phi\kz$ and $\ks\ks\ks$ candidates.
The $\mb$ distribution for the $\bz\to\jpsi\ks$ candidates
and $\pbstar$ distribution for the $\bz\to\jpsi\kl$ candidates are
shown in Fig.~\ref{fig:recjpsi}.
The signal yield for each mode is obtained from an unbinned maximum-likelihood fit to these 
distributions; the $\dE$ distribution is also included in the fit for the
modes without a $\kl$ meson. 
The signal shape for each decay mode is determined from
MC events.
The background has two components: continuum, 
which is modeled using events outside the signal region, and $B\bbar$
background, which is modeled with MC events.
The signal yields are determined to be 
$\Nsigphiks$  for  $\bz\to\phi\ks$,
$\Nsigphikl$  for  $\bz\to\phi\kl$,
$\Nsigetapks$ for $\bz\to\etap\ks$,
$\Nsigetapkl$ for $\bz\to\etap\kl$,
$\Nsigksksks$ for $\bz\to\ks\ks\ks$,
$\Nsigjpsiks$ for $\bz\to\jpsi\ks$ and
$\Nsigjpsikl$ for $\bz\to\jpsi\kl$,
where errors are statistical only.

We determine $\cals$ and $\cala$ for each mode by performing an unbinned
maximum-likelihood fit to the observed $\Dt$ distribution.
The probability density function (PDF) for the signal
distribution, ${\cal P}_{\rm sig}(\Dt;\cals,\cala,\fq,w_l,\dwl)$, 
is given by Eq.~(\ref{eq:psig}) 
fixing $\tau_\bz$ and $\dmd$ at
their world average values~\cite{bib:PDG2006}
and incorporating the effect of incorrect flavor assignment.
The distribution is convolved with the proper-time interval resolution
function $\rsig$, which takes into account the finite vertex
resolution~\cite{bib:resol_nim}.
We determine the following likelihood for each event:
\begin{eqnarray}
P_i
&=& (1-\fol)
\sum_{k}f_k \int\left[
{\cal P}_k (\Dt')R_k (\Dt_i-\Dt')\right] d(\Dt') \nonumber \\
&+&\fol P_{\rm ol}(\Dt_i),
\label{eq:likelihood}
\end{eqnarray}
where $k$ denotes signal, continuum and $B\bbar$ background components.
In the $\bz\to\ks\ks\ks$ and $\jpsi\ks$ samples the $B\bbar$ component
is negligibly small and not included in the fit.
The fraction of each component $f_k$ depends on the $r$ region and
is calculated on an event-by-event basis
as a function of the following variables:
$\dE$ and $\mb$ for $\bz\to\jpsi\ks$;
$\pbstar$ for $\bz\to\jpsi\kl$;
$\pbstar$ and $\rsigbkg$ for $\bz\to\etap\kl$ and $\phi\kl$;
$\mb$ and $\rsigbkg$ for $\bz\to\phi(\to\ks\kl)\ks$;
$\dE$, $\mb$ and $\rsigbkg$ for the other modes.
The PDF ${\cal P}_k(\Dt)$ for background events
is convolved with the resolution function $R_k$ for the
background~\cite{Chen:2005dr,bib:BELLE-CONF-0436}.
The term $P_{\rm ol}(\Dt)$ is a broad Gaussian function that represents
a small outlier component with a fraction of $\fol$~\cite{bib:resol_nim}.
The only free parameters in the fits
are $\cals$ and $\cala$, which are determined by maximizing the
likelihood function
$L = \prod_iP_i(\Dt_i;\cals,\cala)$
where the product is over all events.

\begin{figure}
\includegraphics[width=\figWidth\textwidth]{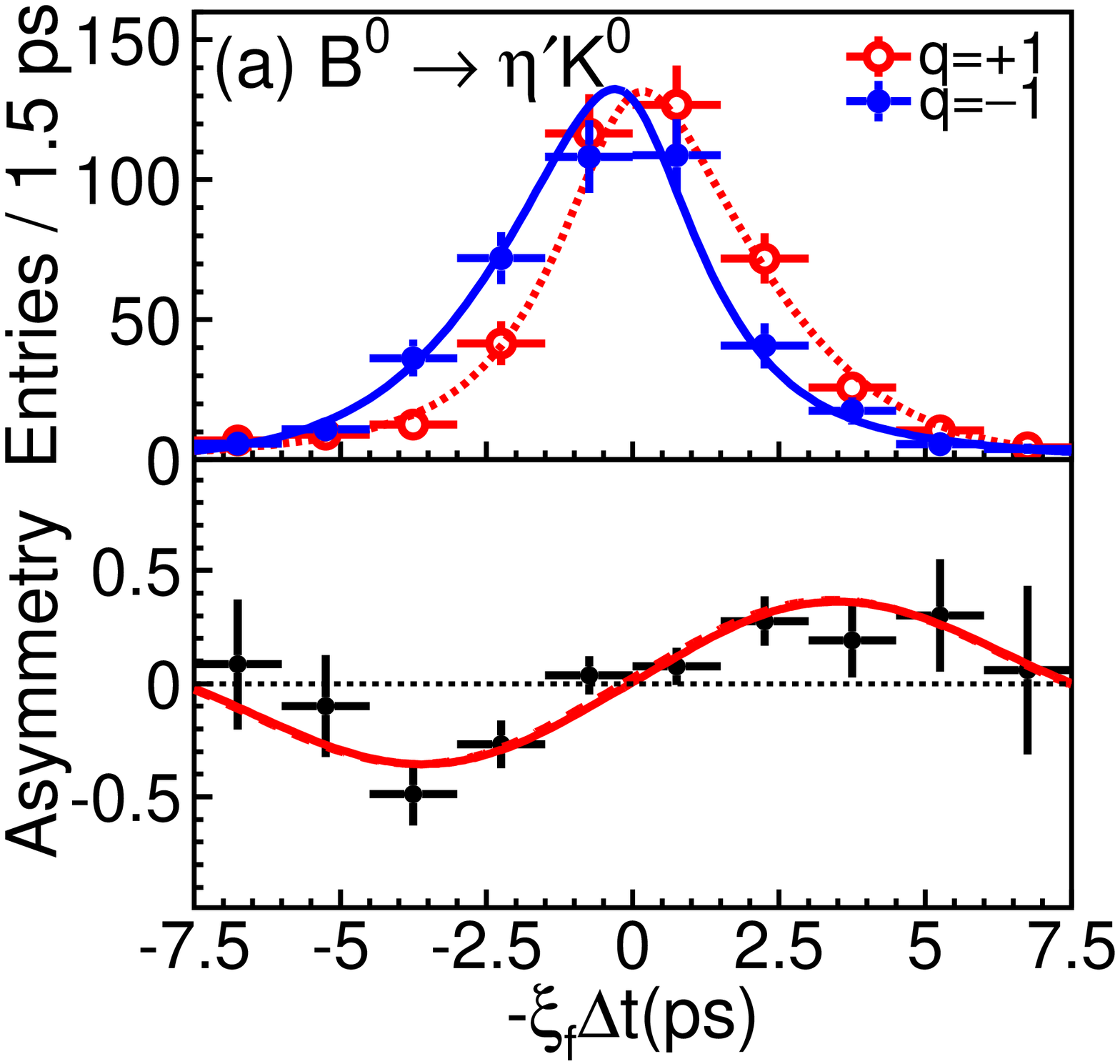}
\includegraphics[width=\figWidth\textwidth]{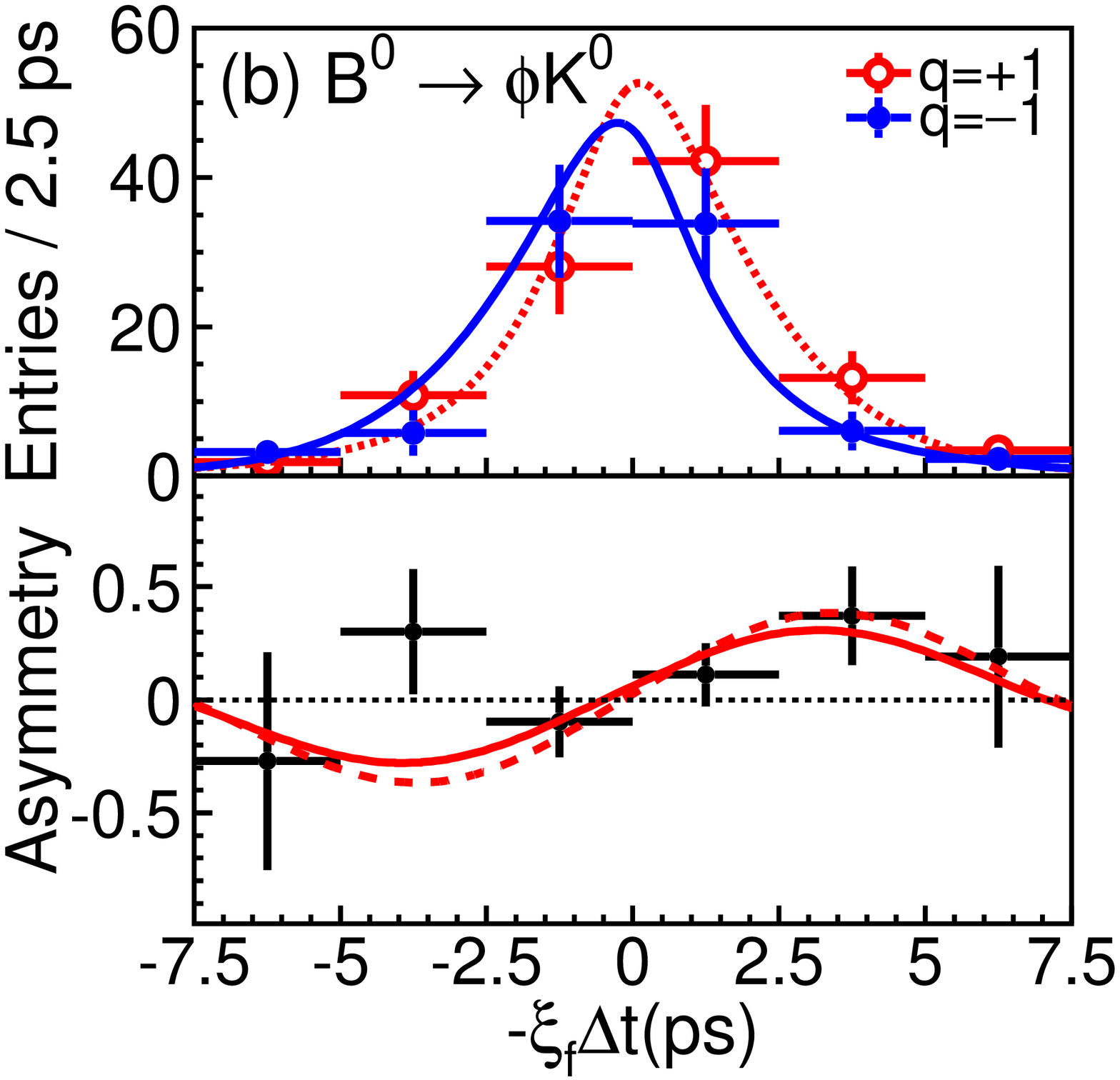}
\includegraphics[width=\figWidth\textwidth]{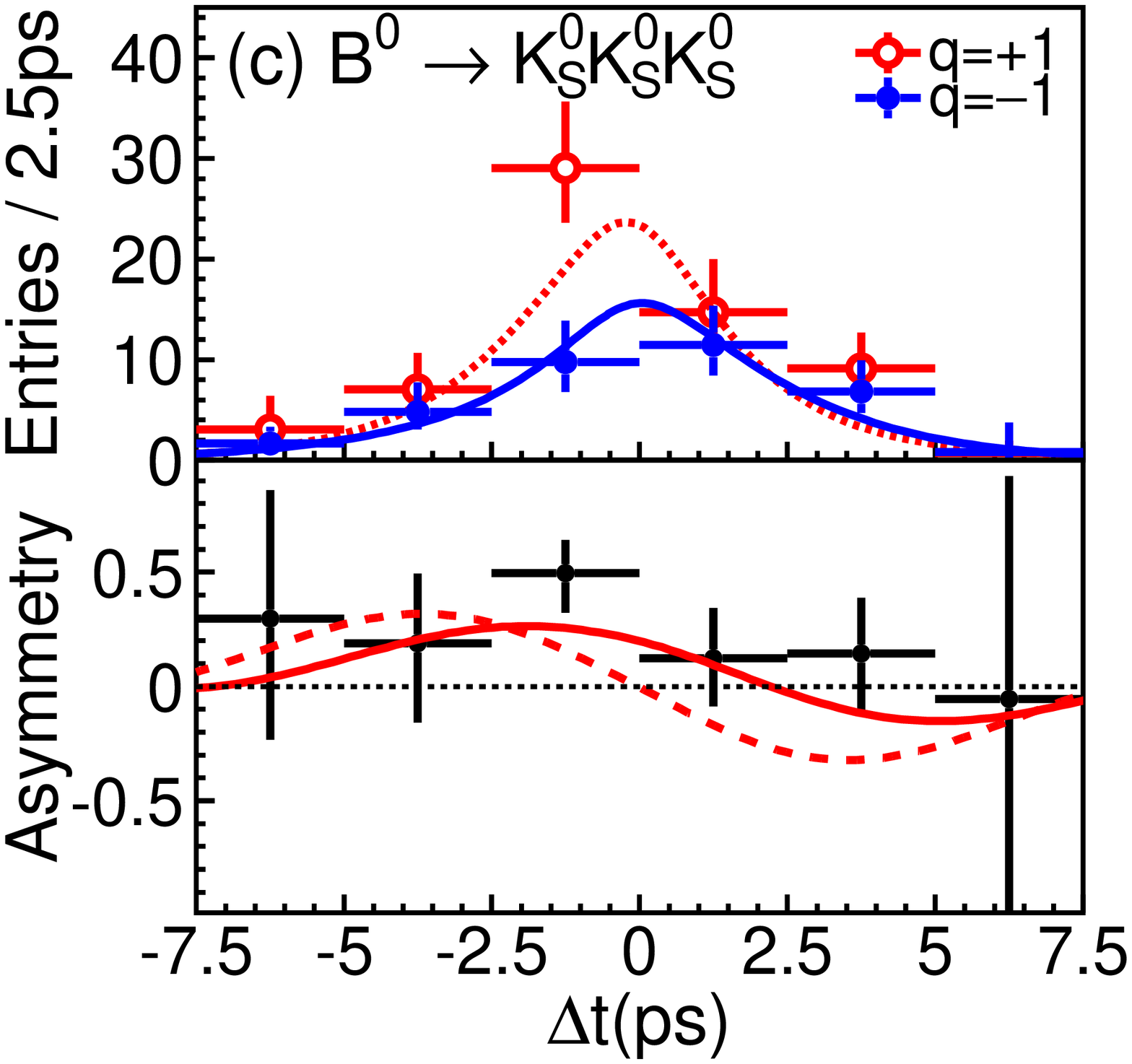}
\includegraphics[width=\figWidth\textwidth]{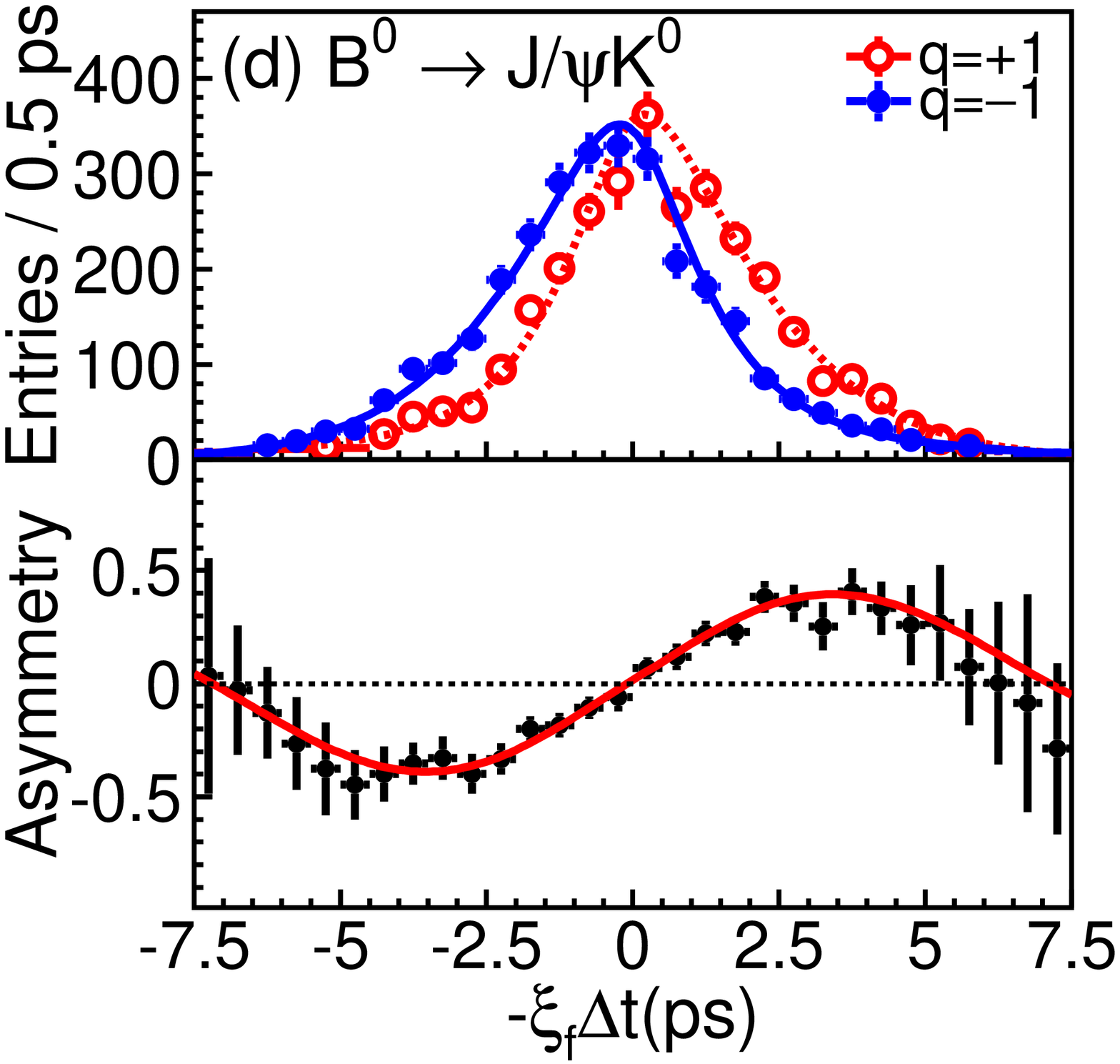}
\caption{
Background subtracted $\Dt$ distributions and asymmetries
for events with good tags ($r>0.5$)
for 
(a) $\bz\to\etap\kz$,
(b) $\bz\to\phi\kz$,
(c) $\bz\to\ks\ks\ks$ 
and
(d) $\bz\to\jpsi\kz$.
In the asymmetry plots, solid curves show the fit results;
dashed curves show the SM expectation from
our $\bz\to\jpsi\kz$ measurement.
}
\label{fig:dtbgsub}
\end{figure}

Table~\ref{tab:result} summarizes
the fit results for $\sinbbeff$ and $\cala$.
These results are consistent with and supersede our previous
measurements~\cite{Chen:2005dr,bib:BELLE-CONF-0436}.
%
\begin{table}[hb]
\caption{Results of the fits to the $\Dt$ distributions.
The first errors are statistical and the second
errors are systematic.
}
\label{tab:result}
\begin{ruledtabular}
\begin{tabular}{cll}
Mode &  \multicolumn{1}{c}{$\sinbbeff$} & \multicolumn{1}{c}{$\cala$}\\ 
\hline
$\phi\kz$   & $\sinbbphikzResult$      & $\AphikzResult$ \\
$\etap\kz$  & $\sinbbetapkzResult$     & $\AetapkzResult$ \\
$\ks\ks\ks$ & $\sinbbksksksResult$     & $\AksksksResult$ \\
$\jpsi\kz$  & $\sinbbjpsikzResult$ & $\AjpsikzResult$ \\
\end{tabular}
\end{ruledtabular}
\end{table}
%
Fits to each individual mode with $\ks$ and $\kl$ yield
(${\cal S}_{\etap\ks}$, ${\cal A}_{\etap\ks}$) = ($\SetapksResultStat$, $\AetapksResultStat$),
(${\cal S}_{\etap\kl}$, ${\cal A}_{\etap\kl}$) = ($\SetapklResultStat$, $\AetapklResultStat$),
(${\cal S}_{\phi\ks}$,  ${\cal A}_{\phi\ks}$)  = ($\SphiksResultStat$,  $\AphiksResultStat$),
(${\cal S}_{\phi\kl}$,  ${\cal A}_{\phi\kl}$)  = ($\SphiklResultStat$,  $\AphiklResultStat$),
(${\cal S}_{\jpsi\ks}$, ${\cal A}_{\jpsi\ks}$) = ($\SjpsiksResultStat$, $\AjpsiksResultStat$)
and
(${\cal S}_{\jpsi\kl}$, ${\cal A}_{\jpsi\kl}$) = ($\SjpsiklResultStat$, $\AjpsiklResultStat$),
where errors are statistical only.
We define the background-subtracted asymmetry in each $\Dt$ bin by $(N_{+}-N_{-})/(N_{+}+N_{-})$,
where $N_{+}$ ($N_{-}$) is the signal yield with $q=+1$ $(-1)$.
Figures~\ref{fig:dtbgsub}(a)-(d) show the $\Dt$ distributions and
asymmetries for good tag quality ($r>0.5$) events.
The sign of each $\Dt$ measurement for final
states with a $\kl$ is inverted
in order to combine results with $\ks$ and $\kl$ mesons. 
%

The dominant sources of systematic error for
$\sinbbeff$ in $\btosqq$ modes come from the uncertainties 
in the resolution function for the signal ($\SetapkzSystResol$ for the $\bz\to\etap\kz$ mode, 
$\SphikzSystResol$ for the $\phi\kz$ mode, $\SksksksSystResol$ for the $\bz\to\ks\ks\ks$ mode)
and 
in the background fraction ($\SetapkzSystBG$, $\SphikzSystBG$, $\SksksksSystBG$).
The effect of $\fzero\kz$ background in the $\phi\kz$ mode (0.02) is 
estimated using the BES measurement of the $\fzero$ lineshape~\cite{bib:BESfzero} and is
included in the background fraction systematic error.
The dominant sources for  $\cala$ in $\btosqq$ modes 
are the effects of tag-side interference~\cite{Long:2003wq} 
($\AetapkzSystTagIntrfr$, $\AphikzSystTagIntrfr$, $\AksksksSystTagIntrfr$),
the uncertainties 
in the background fraction ($\AetapkzSystBG$, $\AphikzSystBG$, $\AksksksSystBG$),
in the vertex reconstruction ($\AetapkzSystVtx$ for all modes) 
and
in the resolution function ($\AetapkzSystResol$, $\AphikzSystResol$, $\AksksksSystResol$).
We study the possible correlations between $\rsigbkg$, $\pbstar$ and $r$ PDFs 
used for $\phi\kl$ and $\etap\kl$, which are neglected in the nominal result, and include
their effect in the systematic uncertainties in the background fraction.
Other contributions come from uncertainties in wrong tag fractions, 
the background $\Dt$ distribution, $\taubz$ and $\dmd$.
A possible fit bias is examined by fitting a large number of MC events
and is found to be small.

The dominant sources of systematic errors for the $\bz\to\jpsi\kz$ mode are
the uncertainties 
in the vertex reconstruction ($\SjpsikzSystVtx$ for $\sinbb$, $\AjpsikzSystVtx$ for $\cala$),
in the resolution function  for the signal ($\SjpsikzSystResol$, $\AjpsikzSystResol$),
in the background fraction ($\SjpsikzSystBG$, $\AjpsikzSystBG$), 
in the flavor tagging ($\SjpsikzSystFbtg$, $\AjpsikzSystFbtg$),
a possible fit bias ($\SjpsikzSystFit$, $\AjpsikzSystFit$) and
the effect of the tag-side interference ($\SjpsikzSystTagIntrfr$, $\AjpsikzSystTagIntrfr$).
Other contributions amount to less than 0.001.
We add each contribution in quadrature to obtain the total systematic uncertainty.

For the $\bz\to\etap\kz$ mode, we observe $CP$ violation
with a significance equivalent to $\SIGMAetapkzAS$ standard deviations
for a Gaussian error, where the significance is calculated using 
the Feldman-Cousins frequentist approach~\cite{bib:FC}. 
The results for $\bz\to\etap\kz$, $\phi\kz$ and $\ks\ks\ks$ decays
are all consistent with the value of $\sinbb$ obtained from the decay $\bz\to\jpsi\kz$
within two standard deviations. No direct $CP$ violation is observed in these decay modes.
Further measurements with much larger data samples are required to search for
new, beyond the SM, $CP$-violating phases in the $b\to s$ transition.


We thank the KEKB group for excellent operation of the
accelerator, the KEK cryogenics group for efficient solenoid
operations, and the KEK computer group and
the NII for valuable computing and Super-SINET network
support.  We acknowledge support from MEXT and JSPS (Japan);
ARC and DEST (Australia); NSFC and KIP of CAS (China); 
DST (India); MOEHRD, KOSEF and KRF (Korea); 
KBN (Poland); MIST (Russia); ARRS (Slovenia); SNSF (Switzerland); 
NSC and MOE (Taiwan); and DOE (USA).



\begin{thebibliography}{999}


\bibitem{bib:lucy}
Y.~Grossman and M.~P.~Worah, 
Phys.\ Lett. B \textbf{395}, 241 (1997);
D.~London and A.~Soni,
Phys.\ Lett. B \textbf{407}, 61 (1997);
T.~Moroi,
Phys.\ Lett.\ B {\bf 493}, 366 (2000);
D.~Chang, A.~Masiero and H.~Murayama,
Phys.\ Rev.\ D {\bf 67}, 075013 (2003);
S.~Baek, T.~Goto, Y.~Okada and K.~Okumura,
Phys.\ Rev.\ D {\bf 64}, 095001 (2001).

\bibitem{bib:sanda}
A.~B.~Carter and A.~I.~Sanda, Phys.\ Rev.\ D \textbf{23}, 1567 (1981);
I.~I.~Bigi and A.~I.~Sanda, Nucl.\ Phys. B \textbf{193}, 85 (1981).

\bibitem{Kobayashi:1973fv}
M.~Kobayashi and T.~Maskawa, Prog.\ Theor.\ Phys. {\bf 49}, 652 (1973).

\bibitem{Cabibbo}
N.~Cabibbo, Phys.\ Rev.\ Lett. {\bf 10}, 531 (1963).
    
\bibitem{footnote:phi1}
Another naming convention $\beta(=\phi_1)$ is also used in literatures.

\bibitem{bib:b2s_SM_uncertainties}
M.~Beneke and M.~Neubert, Nucl.\ Phys. B \textbf{675}, 333 (2003);
M.~Beneke, Phys.\ Lett.\ B {\bf 620}, 143 (2005);
H.-Y.~Cheng, C.-K.~Chua and A.~Soni,
Phys.\ Rev.\ D {\bf 72}, 014006 (2005);
Phys.\ Rev.\ D {\bf 72}, 094003 (2005);
A.~R.~Williamson and J.~Zupan, Phys.\ Rev.\ D {\bf 74}, 014003 (2006);
M.~Gronau, J.~L.~Rosner and J.~Zupan, Phys.\ Rev.\ D {\bf 74}, 093003 (2006).

\bibitem{Chen:2005dr}
Belle Collaboration, K.~F.~Chen {\it et al.},
Phys.\ Rev.\ D {\bf 72}, 012004 (2005).
Belle Collaboration, K.~Sumisawa {\it et al.},
Phys.\ Rev.\ Lett.\ {\bf 95}, 061801 (2005).

\bibitem{bib:BaBar_sss}
BaBar Collaboration, B.~Aubert {\it et al.},
Phys.\ Rev.\ D {\bf 71}, 091102(R) (2005);   
Phys.\ Rev.\ Lett.\ {\bf 94}, 191802 (2005); 
Phys.\ Rev.\ Lett.\  {\bf 95}, 011801 (2005); 

\bibitem{bib:BaBar_new}
BaBar Collaboration, B.~Aubert {\it et al.}, 
Phys.\ Rev.\ Lett. {\bf 98}, 031801 (2007) 

\bibitem{bib:BELLE-CONF-0436}
Belle Collaboration, K.~Abe \textit{et al.},
Phys.\ Rev.\ D \textbf{71}, 072003 (2005).
%






\bibitem{bib:KEKB}
S.~Kurokawa and E.~Kikutani,
Nucl.\ Instrum.\ Methods Phys. Res., Sect. A {\bf 499}, 1 (2003),
and other papers included in this volume.

\bibitem{Belle}
Belle Collaboration, A.~Abashian {\it et al.},
Nucl.\ Instrum.\ Methods Phys. Res., Sect. A {\bf 479}, 117 (2002);
Y. Ushiroda (Belle SVD2 Group),
Nucl.\ Instrum.\ Methods Phys. Res., Sect. A {\bf 511}, 6 (2003).

\bibitem{bib:b2s_lp05}
Belle Collaboration, K.~Abe {\it et al.},
hep-ex/0507037

\bibitem{Garmash:2003er}
Belle Collaboration, A.~Garmash {\it et al.},  
Phys.\ Rev.\ D {\bf 69}, 012001 (2004);
Phys.\ Rev.\ D {\bf 71}, 092003 (2005).

\bibitem{bib:fbtg_nim}
H.~Kakuno {\it et al.},
Nucl.\ Instrum.\ Methods Phys. Res., Sect. A {\bf 533}, 516 (2004).

\bibitem{bib:resol_nim}
H.~Tajima {\it et al.},
Nucl.\ Instrum.\ Methods Phys. Res., Sect. A {\bf 533}, 370 (2004).


\bibitem{bib:PDG2006}
W.-M. Yao {\it et al.}, J. Phys. G {\bf 33}, 1(2006).

\bibitem{bib:BESfzero}
BES Collaboration, M.~Ablikim {\it et al.},
Phys.\ Lett.\ B {\bf 607}, 243 (2005).
    
\bibitem{Long:2003wq}
O.~Long, M.~Baak, R.~N.~Cahn and D.~Kirkby,
Phys.\ Rev.\ D {\bf 68}, 034010 (2003).


\bibitem{bib:FC}
G.~J.~Feldman and R.~D.~Cousins,
Phys.\ Rev.\ D {\bf 57}, 3873 (1998).

\end{thebibliography}
\end{document}